\documentstyle[12pt,aasms4]{article}
\lefthead{Meyer, Jura, \& Cardelli}
\righthead{Abundance of Interstellar Oxygen}

\begin{document}

\title{The Definitive Abundance of Interstellar Oxygen\altaffilmark{1}}

\author{David M. Meyer}
\affil{Department of Physics and Astronomy, Northwestern University,
Evanston, IL  60208; meyer@elvis.astro.nwu.edu}
\authoremail{meyer@elvis.astro.nwu.edu}

\author{M. Jura}
\affil{Department of Physics and Astronomy, University of California,
Los Angeles, CA  90095-1562; jura@clotho.astro.ucla.edu}
\authoremail{jura@clotho.astro.ucla.edu}

\and

\author{Jason A. Cardelli\altaffilmark{2}}
\affil{Department of Astronomy and Astrophysics, Villanova University,
Villanova, PA  19085}

\altaffiltext{1}{Based on observations obtained with the NASA/ESA
{\it Hubble Space Telescope} through the Space Telescope Science
Institute, which is operated by the Association of Universities for
Research in Astronomy, Inc., under NASA contract NASA-26555.}
\altaffiltext{2}{Deceased 1996 May 14.}

\begin{abstract}
Using the Goddard High Resolution Spectrograph (GHRS) onboard the
{\it Hubble Space Telescope}, we have obtained high S/N
echelle observations of the weak interstellar O I $\lambda$1356
absorption toward the stars $\gamma$ Cas, $\epsilon$ Per, $\delta$ Ori,
$\epsilon$ Ori, 15 Mon, $\tau$ CMa, and $\gamma$ Ara.  In combination
with previous GHRS measurements in six other sightlines ($\zeta$ Per,
$\xi$ Per, $\lambda$ Ori, $\iota$ Ori, $\kappa$ Ori, and $\zeta$ Oph),
these new observations yield a mean interstellar gas-phase oxygen
abundance (per 10$^6$ H atoms) of 10$^6$ O/H $=$ 319 $\pm$ 14.
The largest deviation from the mean is less than 18\%, and
there are no statistically significant variations in the measured O
abundances from sightline to sightline and no evidence of
density-dependent oxygen depletion from the gas phase.  Assuming
various mixtures of silicates and oxides, the abundance of
interstellar oxygen tied up in dust grains is unlikely to surpass
10$^6$ O/H $\approx$ 180.
Consequently, the GHRS observations imply
that the {\it total} abundance of interstellar oxygen (gas plus
grains) is homogeneous in the vicinity of the Sun 
and about 2/3 of the solar value
of 10$^6$ O/H $=$ 741 $\pm$ 130.  This oxygen deficit is consistent
with that observed in nearby B stars and similar to that recently
found for interstellar krypton with GHRS.  Possible explanations
for this deficit include: (1) early solar
system enrichment by a local supernova, (2) a recent infall of
metal-poor gas in the local Milky Way, or (3) an outward diffusion of
the Sun from a smaller galactocentric distance.
\end{abstract}


\section{Introduction}

Oxygen is the most abundant element in the Galaxy
after hydrogen and helium.  Consequently, it is important to establish
the current-epoch O abundance accurately
for studies of Galactic chemical evolution (\cite{tim95}).  A
considerable effort has recently gone into determining the
abundances of oxygen and other elements in nearby B stars since these
young stars should most closely reflect the current ISM abundance
pattern (\cite{gie92,kil92,cun94,kil94}).  These studies yield a
median B-star oxygen abundance (per 10$^6$ H atoms) of
10$^6$ O/H $\approx$ 450 which is about 2/3 of the Grevesse \& Noels
(1993) solar value (10$^6$ O/H $=$ 741 $\pm$ 130).  This result is
inconsistent with the traditional assumptions that the solar
abundance reflects that of the ISM at the time of the
Sun's formation 4.6 Gyr ago and that the interstellar O abundance should
increase slowly over time (\cite{aud76}, Timmes et al.\ 1995).

A simple interpretation of the conflict between the solar and B star
abundances is that it is a manifestation of the abundance scatter
in their respective stellar populations.  For example, Cunha \&
Lambert (1994) find a spread of $\pm$0.2 dex among the oxygen
abundances in their sample of Orion association B stars.  Also,
in a sample of F and G stars of similar age and Galactocentric
radius to the Sun, \cite{edv93} find an iron abundance spread of
$\pm$0.25 dex with the Sun among the most metal-rich cases.  Such
scatter in the stellar abundances is strongly suggestive of localized
abundance inhomogeneities in the ISM\@.  Yet, as discussed by
\cite{roy95}, a variety of hydrodynamical processes operating
in the Galactic disk should keep the gas chemically well-mixed
on short time scales.  In order to determine the homogeneity of the
local interstellar medium at a level capable of distinguishing
between a solar and a B star oxygen abundance, it is necessary to
obtain very sensitive observations.

Since O I (13.618 eV) and H I (13.598 eV) have nearly the same
ionization potentials, O I is the dominant form of gaseous oxygen
in diffuse interstellar H I clouds.  Although the prominent O I
$\lambda$1302 absorption line is typically saturated in diffuse
sightlines, the very weak intersystem transition at 1355.598 \AA\
can yield accurate gas-phase O abundances if measured with
sufficient sensitivity.  Observations of this line with the
{\it Copernicus} satellite indicate a mean interstellar gas-phase
oxygen abundance that is 40\% to 70\% of the solar value
(\cite{yor83,kee85}).  However, the scatter in these data is too
great to rule out a solar abundance of interstellar oxygen
especially since some of the O is tied up in dust grains.  This
scatter is primarily due to the uncertainties associated with
measuring the weak O I $\lambda$1356 line strengths.

The ability of the Goddard High Resolution Spectrograph (GHRS)
onboard the {\it Hubble Space Telescope} ({\it HST}) to obtain
UV spectra with higher resolution and signal-to-noise (S/N) ratios
than those acquired with {\it Copernicus} has made it possible to take
interstellar abundance studies another step forward with accurate
observations of very weak absorption lines (\cite{sav96}).  Utilizing
high S/N GHRS observations of the O I $\lambda$1356 absorption in the
low density sightlines toward the stars $\iota$ Ori and $\kappa$ Ori,
\cite{mey94} have found a total oxygen abundance (gas plus grains) in
Orion that is consistent with the stellar (\cite{cun94}) and nebular
determinations (\cite{bal91,rub91,ost92,pei93}).
In this paper, we present new O I $\lambda$1356 data;
our total GHRS sample includes 13 sightlines toward
stars in seven distinctly different Galactic directions at distances
ranging from 130 to 1500 pc, with most closer than 500 pc.  
These sightlines were particularly
chosen for their wide range in physical conditions so as to search
for evidence of density-dependent depletion variations in the
gas-phase oxygen abundance as well as spatial variations.

\section{Observations}

The GHRS observations of the interstellar O I $\lambda$1356 absorption
line toward the stars $\gamma$ Cas, $\epsilon$ Per, $\delta$ Ori,
$\epsilon$ Ori, 15 Mon, $\tau$ CMa, and $\gamma$ Ara were obtained
in 1995 October and November using the echelle-A grating and the
2\farcs0 large science aperture.  The detailed characteristics
and in-flight performance of the GHRS instrument are discussed
by \cite{bra94} and \cite{hea95}.  The observations of each star
consisted of multiple FP-Split exposures centered near 1356 \AA.
An FP-Split breaks up an exposure into four subexposures taken at
slightly different grating positions so as to better characterize
and minimize the impact of the GHRS Digicon detector's fixed pattern
noise on the S/N ratio of the data.  Each of these subexposures
was sampled two or four times per Digicon diode (depending on the
brightness of the star) at a velocity resolution of 3.5 km s$^{-1}$.

The data reduction procedure discussed in detail by \cite{car94a}
was utilized to maximize the S/N ratio of the O I spectra.  Basically,
this process involves four steps: (1) the subexposures comprising
each FP-Split exposure are merged in diode space so as to create
a template of the fixed pattern noise spectrum, (2) each subexposure
is divided by this noise template, (3) all of the rectified
subexposures are aligned in wavelength space using the interstellar
lines as a guide, and (4) the aligned subexposures are summed to
produce the net O I spectrum of each star.  As illustrated in Figure 1
for five of the stars in our echelle sample, the resulting
continuum-flattened spectra reveal convincing detections
of the interstellar O I $\lambda$1356 line in all seven sightlines.
The measured S/N ratios of these spectra are all in the 400 to 600
range.

The GHRS spectra of $\lambda$ Ori and $\zeta$ Per displayed in
Figure 1 were obtained in 1994 February using the
G160M grating and the 0\farcs25 small science aperture.  These
spectra were reduced in the same manner as the echelle data and
are each characterized by a velocity resolution of 16 km s$^{-1}$ and
a S/N ratio of about 500.  The measured equivalent widths of the
interstellar O I $\lambda$1356 absorption in these spectra as well
as those in the echelle sightlines are listed in Table 1.  The
uncertainties in these line strengths reflect the statistical
and continuum placement errors summed in quadrature.  Table 1 also
includes the previously reported GHRS O I $\lambda$1356 measurements
toward $\iota$ Ori and $\kappa$ Ori (Meyer et al.\ 1994),
$\xi$ Per (\cite{car91}), and $\zeta$ Oph (\cite{sav92}).

The O I column densities listed in Table 1 were calculated from
the $\lambda$1356 equivalent widths using the Zeippen, Seaton, \&
Morton (1977) oscillator strength of $f=1.248\times10^{-6}$.
Although the quoted uncertainty in this $f$-value is 15\%,
\cite{sof94} have empirically verified that it is consistent with the
better-determined $f$-value appropriate for the O I $\lambda$1302
transition.  In the cases of $\gamma$ Cas, $\epsilon$ Per,
$\delta$ Ori, $\iota$ Ori, $\epsilon$ Ori,
$\kappa$ Ori, 15 Mon, and $\tau$ CMa, the $\lambda$1356
absorption is weak enough for $N$(O I) to be confidently derived
under the assumption that the line is optically thin.  In the cases
of $\zeta$ Per, $\lambda$ Ori, and $\gamma$ Ara, a slight correction
for saturation was applied using a Gaussian curve-of-growth with
respective $b$-values of 2.0$^{+2.0}_{-0.5}$, 5.0$^{+\infty}_{-2.5}$,
and 3.0$^{+\infty}_{-1.5}$ km s$^{-1}$.  These $b$-values were
estimated from GHRS observations of the interstellar Mg II
$\lambda\lambda$1239.9,1240.4 doublet toward $\zeta$ Per, the
Mg II and N I $\lambda\lambda$1160,1161 (\cite{mey97})
doublets toward $\lambda$ Ori, and the O I $\lambda$1356 line
width toward $\gamma$ Ara.  The resultant O I column
densities for $\zeta$ Per, $\lambda$ Ori, and $\gamma$ Ara
are 24\%, 4\%, and 6\% greater than their weak line limits,
respectively.  The $N$(O I) values corresponding to the slightly
saturated $\lambda$1356 lines toward $\xi$ Per and $\zeta$ Oph were
taken from the detailed analyses of these sightlines by Cardelli et
al.\ (1991) and Savage et al.\ (1992).  The uncertainties in the O I
column densities listed in Table 1 reflect the estimated errors in
the $\lambda$1356 equivalent width measurements and the saturation
corrections (where applied).  These errors do not include
the uncertainty in the $\lambda$1356 $f$-value because it would
affect all of the column densities in the same way.

\section{Results}

In order to put the GHRS oxygen results in perspective, it is
instructive to compare and analyze them in concert
with the best {\it Copernicus} satellite observations
of the interstellar O I $\lambda$1356 line.  Table 2 lists
the 14 sightlines toward which the equivalent width of this line
has been measured at the 4$\sigma$ level or better with
{\it Copernicus} (\cite{boh83}, Zeippen et al.\ 1977).  Among the four
sightlines in common between the GHRS and {\it Copernicus} samples,
$\zeta$ Oph yields similar O I line strengths while the
$\epsilon$ Per, $\lambda$ Ori, and $\kappa$ Ori lines are weaker
in the more sensitive GHRS spectra.
In deriving the {\it Copernicus} O I column densities listed in
Table 2, $\epsilon$ Per and $\kappa$ Ori were assumed to be optically
thin while the $\lambda$ Ori and $\zeta$ Oph lines were corrected
for saturation in the same way as the GHRS data.  Since the other
sightlines in the {\it Copernicus} sample have appreciably stronger
O I lines, saturation is more of a concern than in the general case
of the GHRS sample.  For these sightlines, saturation corrections
were applied using a single-component Gaussian curve-of-growth and
the $b$-value estimates listed in Table 2. The $b$-value estimates are
based in part on {\it Copernicus} observations of interstellar Cl I and
P II in these sightlines (\cite{jen86}). The impact of the saturation
corrections on $N$(O I) ranges from 12\% over the weak-line limit for
$\delta$ Sco to 68\% for $\rho$ Oph, and less than 30\% for most of the
other sightlines.  The quoted errors in the derived O I column
densities reflect the uncertainities in these corrections as well
as those in the $\lambda$1356 line strength measurements.

The total hydrogen column densities ($N$(H) $=$ 2$N$(H$_2$) $+$ $N$(H I))
listed in Tables 1 and 2 were determined in the same manner for each
sightline in the GHRS and {\it Copernicus} samples.  These values
reflect the H$_2$ column densities measured by Savage et al.\ (1977)
and the weighted means of the Bohlin, Savage, \& Drake (1978) and
Diplas \& Savage (1994) $N$(H I) data.  The uncertainties in the resulting
oxygen abundances (per 10$^6$ H atoms) in Tables 1 and 2 reflect the
propagated errors in both $N$(O I) and $N$(H).  Taken together, the
GHRS sightlines yield a weighted mean interstellar gas-phase oxygen
abundance of 10$^6$ O/H $=$ 319 $\pm$ 14 while the {\it Copernicus}
data yields 10$^6$ O/H $=$ 361 $\pm$ 20.  Although the {\it Copernicus}
mean is heavily weighted by the accurate value toward $\zeta$ Oph,
both samples are indicative of an interstellar gas-phase O abundance
that is appreciably below the solar value of
10$^6$ O/H $=$ 741 $\pm$ 130 (Grevesse \& Noels 1993).
The key improvement of the GHRS data over the {\it Copernicus} data is
the greater accuracy of the individual GHRS measurements
(especially of the weaker O I lines).  In the GHRS data,
the largest deviation of O/H
from the mean value is 18\% compared to a range in the
{\it Copernicus} data of a factor of 3.

As reviewed by Jenkins (1987), the interstellar gas-phase abundances
of many elements measured by {\it Copernicus} decrease as a function of
the mean sightline hydrogen density, $n_H$ $=$ $N$(H)/$r$, where $r$
is the distance to the background star.
These elemental depletions from the gas phase reflect both the growth
of dust grains in denser clouds and grain destruction in more diffuse
environments.  Using the stellar distances listed in Diplas \&
Savage (1994), we have calculated $n_H$ for each of the sightlines
in our GHRS and {\it Copernicus} samples and plotted them versus the
corresponding oxygen abundances in Figure 3.  As might be expected
from Figure 2, there is no significant evidence of variations in
the gas-phase O abundance as a function of $n_H$ in either sample.
Although the {\it Copernicus} data samples more denser sightlines,
the GHRS data pins down the oxygen gas abundance in the most
diffuse clouds at a level that is completely consistent with the
higher density cases.
The absence of abundance
variations as a function of $n_H$ in the GHRS data suggests that
there is negligible exchange between gas and dust in these diffuse
sightlines and that the total (gas plus dust) abundance of oxygen
must not vary significantly in the local ISM\@.

A better barometer of diffuse cloud conditions
is the fractional abundance of molecular hydrogen,
$f$(H$_2$) $=$ 2$N$(H$_2$)/$N$(H) (\cite{car94b}).
Sightlines separate rather distinctly into groups
with low and high $f$(H$_2$) values due to the difference between
UV transparent and H$_2$ self-shielding environments.  Even for
weakly depleted elements like Ge (\cite{car94b}) and Zn
(\cite{rot95,sem95}), the gas-phase abundances are higher in the
low $f$(H$_2$) group than in the high group signifying both the
presence of dust grains and changes in the elemental dust abundance
due to interstellar grain growth and/or destruction.  Figure 4
clearly shows that the interstellar gas-phase O abundances measured with
GHRS are both well-sampled as a function of $f$(H$_2$) and exhibit
no dependence on this parameter.  Indeed, the mean abundance in the
7 sightlines with log $f$(H$_2$) $<$ -2.0 (10$^6$ O/H $=$ 325 $\pm$ 20)
is essentially the same as that in the 6 sightlines with
log $f$(H$_2$) $>$ -2.0 (10$^6$ O/H $=$ 312 $\pm$ 19).  Thus, any
significant reservoir of interstellar oxygen in diffuse clouds
other than the atomic gas must be resilient enough to survive in a variety
of environments.

\section{Discussion}

Under the traditional assumption that the {\it cosmic} elemental
abundances reflect those in the solar system, the mean gas-phase
abundance of interstellar oxygen measured by GHRS implies an
O dust fraction of 10$^6$ O/H $\approx$ 420.  However, it has been
known for some time that an elemental inventory of the likely
constituents of interstellar dust yields appreciably less solid-state
oxygen than this inferred amount (\cite{gre74,mey89}).
In particular, assuming various mixtures
of oxygen-bearing grain compounds such as the silicates
pyroxene [(Mg,Fe)SiO$_3$] and olivine [(Mg,Fe)$_2$SiO$_4$] and oxides
like Fe$_2$O$_3$, it is difficult to increase the O dust fraction
much beyond 10$^6$ O/H $\approx$ 180 (Cardelli et al.\ 1996) simply
because the requisite metals are far less abundant than oxygen
([O:Si:Mg:Fe]$_{solar}$ $\approx$ [24:1:1:1]).  If these metals have
total (gas plus dust) underabundances similar to that derived for
oxygen, the implied O dust fraction would be
10$^6$ O/H $\approx$ 120 instead of 10$^6$ O/H $\approx$ 180.  It would
be hard to hide a significant amount in molecules like CO or O$_2$
in the diffuse sightlines observed by GHRS without leaving any
trace of O abundance variations as a function of $f$(H$_2$).  For
similar reasons, the ``missing'' oxygen is unlikely to be locked up
in icy grain mantles or ice grains (\cite{gre74}).  Such carriers
would also leave unmistakable signatures like the 3.07 $\mu$m O-H
stretch ``ice'' feature that are not observed in diffuse
sightlines (\cite{whi88}).  Thus, unless there is some other resilient
form of oxygen in the diffuse ISM, it appears clear from our GHRS
observations that the {\it total} abundance of interstellar O is
about 2/3 of the solar value.  This result is consistent with the
conclusions of the previous GHRS oxygen studies (Meyer et al.\ 1994,
Cardelli et al.\ 1996) that used
subsets of the complete sightline sample presented here.

The possibility that the interstellar oxygen measurements are sampling
an overall deficit in local ISM elemental abundances has been
enhanced by recent GHRS observations of interstellar krypton.  Based
on measurements in ten sightlines, \cite{car97} find a mean
interstellar gas-phase Kr abundance that is about 60\% of the solar
system abundance.  Since Kr, as a noble gas, should not be depleted
much into dust grains, this gas-phase abundance reflects a true
interstellar deficit similar to what we find for oxygen.
Furthermore, the abundance of krypton, like oxygen, is remarkably
homogeneous from sightline to sightline independent of diffuse
cloud conditions.  This homogeneity is reflected in Figure 5 where
we plot the interstellar gas-phase O/Kr abundance ratio as a function
of $f$(H$_2$) for the GHRS sightlines in common between this study and
that of \cite{car97}.  The current data are consistent
with a picture where the abundances of all of
the elements in the local ISM are generally about 2/3 of their solar
system values (Snow \& Witt 1995, 1996).

The interstellar abundance deficit suggested by the GHRS observations
of O and Kr fortifies the results of nearby B star measurements
of the current epoch abundances of O and other elements
(\cite{gie92,kil92,cun94,kil94}).  As discussed earlier, these
studies yield median B-star CNO abundances that are also about
2/3 of the solar values.
The implication of this result is that something
unusual happened to either the Sun or the local ISM in the context
of standard models of Galactic chemical evolution
(\cite{aud76}, Timmes et al.\ 1995).

The $\pm$0.05 dex spread in the
interstellar oxygen abundances is appreciably less than the
$\pm$0.2 dex oxygen spread in Orion B stars (Cunha \& Lambert 1994)
and the $\pm$0.25 dex Fe abundance spread in the solar-like star
sample of Edvardsson et al.\ (1993).  If one believes that these
stellar abundance spreads are real and not due to observational
error, the question arises as to how to make stars with such large
abundance variations out of a very well-mixed ISM\@.  The GHRS
data certainly makes it difficult now to explain the solar anomaly
simply as the result of a typical ISM abundance fluctuation.  There
are three models to explain why the ISM has a lower oxygen
abundance than does the Sun.

1. Based on isotopic anomalies involving $^{26}$Al and other elements
in meteorites and cosmic rays, the idea that the early solar system
was chemically enriched by a local supernova explosion has long
been popular (\cite{ree78,lee79,oli82}).  At first glance, this
idea would seem to be a reasonable explanation for the overabundance
of oxygen in the Sun.  If the solar system originated in a molecular
cloud with active OB star formation, a first generation of massive
stars could have evolved quickly and enriched the gas in heavy
elements such as oxygen which would later be incorporated in the
Sun.  Cunha \& Lambert (1994) have found evidence of such a process
in the Orion OB association where elements such as oxygen, which
is produced in abundance by Type II supernovae, exhibit larger
abundance spreads in the B stars than do elements like nitrogen.  In the
context of the Edvardsson et al.\ (1993) study of solar-like stars,
if this kind of cloud self-enrichment process were common, it could
also explain the stellar Fe abundance spread as well as the
Sun's position near the top.  Yet, our GHRS data indicates that the
ISM abundance inhomogeneities produced by any such process must
be quickly damped out.  In particular, our five GHRS Orion sightlines
yield a spread of only $\pm$0.05 dex in their interstellar oxygen gas
abundances and a mean value of 10$^6$ O/H $=$ 305 $\pm$ 24 which
is completely consistent with that of the other GHRS sightlines
(10$^6$ O/H $=$ 325 $\pm$ 17).  Even if this mixing problem can
be accommodated through Galactic hydrodynamical processes
(Roy \& Kunth 1995), the supernova enrichment hypothesis still faces
the challenge of creating similar overabundances for a variety of
elements in the Sun.  For example, in addition to O and Kr, the early
GHRS results on interstellar nitrogen also yield a 2/3 solar abundance
for this element (Cardelli et al.\ 1996).  Given the steep
relative yield of O to N in Type II supernovae (\cite{oli82})
as compared to their present-day interstellar abundances,
it is difficult to understand how one or more such explosions
could have produced similar overabundances of these two elements,
let alone others, in the protosolar nebula.

2. As discussed by Meyer et al.\ (1994) and Roy \& Kunth (1995), another
approach to understanding the underabundance of interstellar
oxygen is to invoke a recent infall of metal-poor extragalactic gas
in the local Milky Way.  The idea of infall, whether gradual or
episodic, has long been recognized as a potentially important
component of Galactic chemical evolution
(\cite{aud76,may81,pit89,chi97}).  Recent observations of high-velocity
gas in the Galactic halo indicate that at least some of these infalling
clouds have metallicities as low as 10\% solar (\cite{kun94,lu94}).
\cite{com94} have suggested that the impact of a
$\approx$10$^6$ M$_{\sun}$ extragalactic cloud with the local Milky
Way some 10$^8$ years ago could explain the origin and characteristics
of the nearby early-type stars and molecular clouds that constitute
Gould's Belt.  In terms of the resultant metallicity of the mixed gas,
such a collision could also have diluted the heavy element abundances
in the local ISM below their solar values.  Since this dilution would
affect all of the elements in the same way, the similar interstellar
underabundances observed for O and Kr could easily be explained through
an infall model.  Conversely, such a model would have serious problems
if any element was found not to exhibit this underabundant pattern.
Sulfur is a potential candidate to break this pattern since
\cite{fit97} have measured near-solar interstellar gas-phase S
abundances toward three stars with high-quality GHRS data.  
However, these abundances
are quite uncertain due to the considerable saturation of the S II
$\lambda\lambda$1251,1254,1260 absorption lines and the likely
possibility that a significant fraction of the S II (which has an
ionization potential of 23.3 eV) originates in H II regions.

Another prediction of the local infall hypothesis would be that the
abundances just beyond Gould's Belt should be closer to the solar values.
In terms of the B stars within a kpc or so, the data are generally
inconclusive on this point with O abundance spreads of about
$\pm$0.2 dex and no systematic variations found
(\cite{geh85,fit92,kil94a,kau94,sma96}).  However, in a comprehensive
study of B stars over a large range of galactocentric distance
(6 $\leq$ $R_g$ $\leq$ 18 kpc), \cite{sma97} find an oxygen abundance
gradient of -0.07 $\pm$ 0.01 dex kpc$^{-1}$ that they claim should
be representative of the present-day Galactic ISM\@.  This large-scale
gradient is consistent with that measured for oxygen in H II regions
(\cite{sha83,sim95,aff96}) and planetary nebulae (\cite{mac94}).
Unfortunately, the small-scale scatter in all of these gradient measures
is too large to shed much light on the local infall hypothesis.

3. \cite{wie96}
suggest that the Sun actually formed in the more metal-rich ISM at a
galactocentric distance of $R_g$ $=$ 6.6 $\pm$ 0.9 kpc and has
migrated over the past 4.6 Gyr to its current distance of
$R_g$ $=$ 8.5 kpc.  This scenario is predicated on a very smooth
ISM metallicity gradient and a process of radial stellar diffusion
that would lead to both the Sun's enhanced Fe metallicity and the
observed $\pm$0.25 dex spread in the Fe abundances of nearby
solar-like stars (Edvardsson et al.\ 1993).  In terms of our GHRS
interstellar O abundances, this idea is attractive because
it could explain both the solar oxygen overabundance and how such
a well-mixed local ISM could co-exist with the much greater
metallicity spreads of local stellar populations.  The key challenge
for the Wielen et al.\ scenario is working out the basic mechanism
through which the stellar orbits can appreciably migrate radially.
Furthermore, based on the $\pm$0.2 dex spread of the O abundances
in the Orion B stars (Cunha \& Lambert 1994), it is clear that stellar
diffusion cannot be responsible for every large stellar abundance
spread.

All three of these models are subject to future observational tests.
If the abundances of all of the elements in the local ISM are
2/3 solar, then the model that the Sun was enriched by a local
supernova is untenable.  Accurate measurements of interstellar
abundances at distances greater than 1 kpc will allow us to test
models which predict spatial variations of these quantities.

\acknowledgments

This work was supported by NASA through grant NAG5-2178 to UCLA.

\clearpage

\figcaption[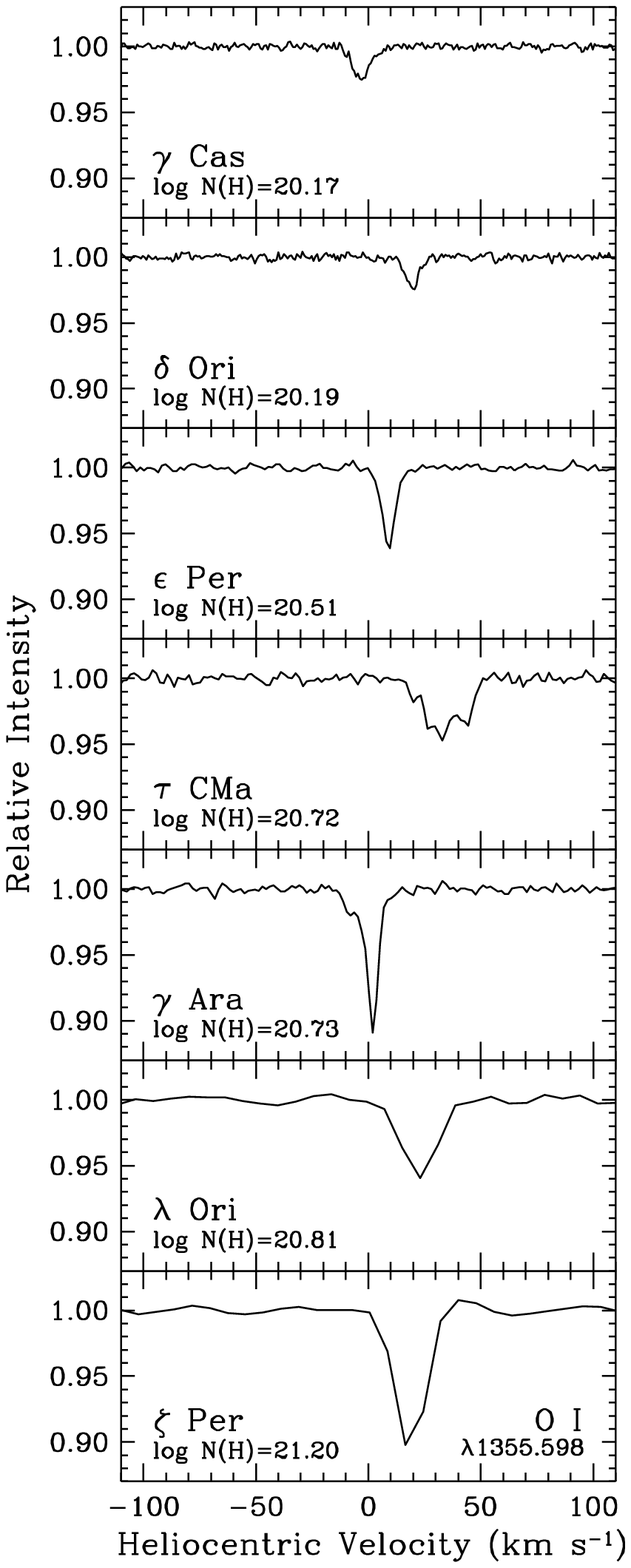]{{\it HST} GHRS spectra of the interstellar O I
$\lambda$1355.598 absorption line toward $\gamma$ Cas, $\delta$ Ori,
$\epsilon$ Per, $\tau$ CMa, and $\gamma$ Ara at a velocity
resolution of 3.5 km s$^{-1}$ and toward $\lambda$ Ori and
$\zeta$ Per at a resolution of 16 km s$^{-1}$.  The spectra are
displayed from top to bottom in order of increasing total
hydrogen column density in the observed sightlines.  The measured
S/N ratios of these spectra are all in the 400 - 600 range.
The measured equivalent widths of the O I lines are listed in
Table 1.}

\figcaption[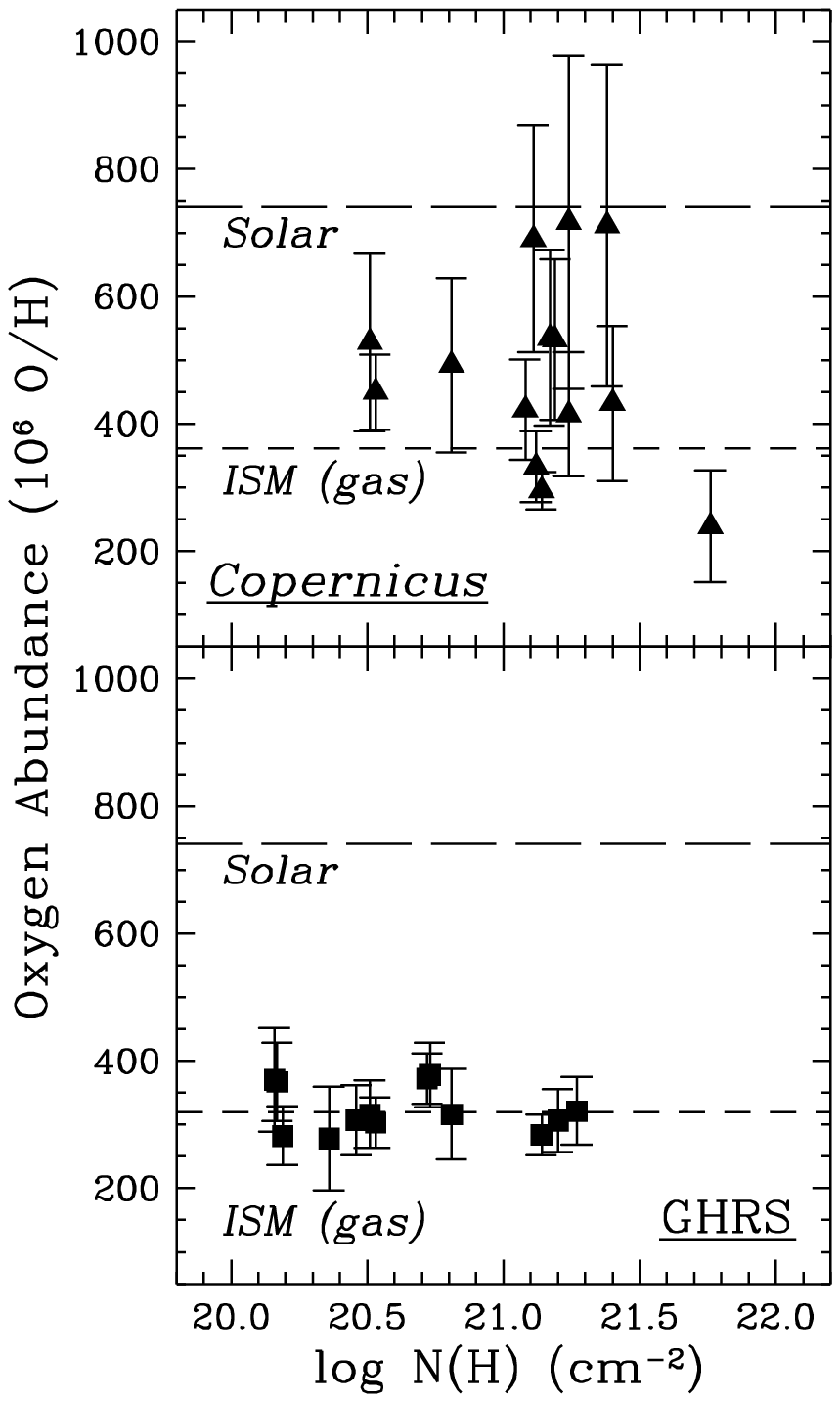]{The interstellar oxygen abundances measured in our
{\it Copernicus} and GHRS samples as a function of the logarithmic
total hydrogen column density, $N$(H) $=$ 2$N$(H$_2$) $+$ $N$(H I),
in the observed sightlines.  The short-dashed lines through the
data points represent the respective {\it Copernicus} and GHRS
weighted mean values (per 10$^6$ H
atoms) of 10$^6$ O/H $=$ 361 $\pm$ 20 and
10$^6$ O/H $=$ 319 $\pm$ 14 for the interstellar gas-phase
oxygen abundance.  The long-dashed lines represent the
Grevesse \& Noels (1993) solar oxygen abundance
(10$^6$ O/H $=$ 741 $\pm$ 130).}

\figcaption[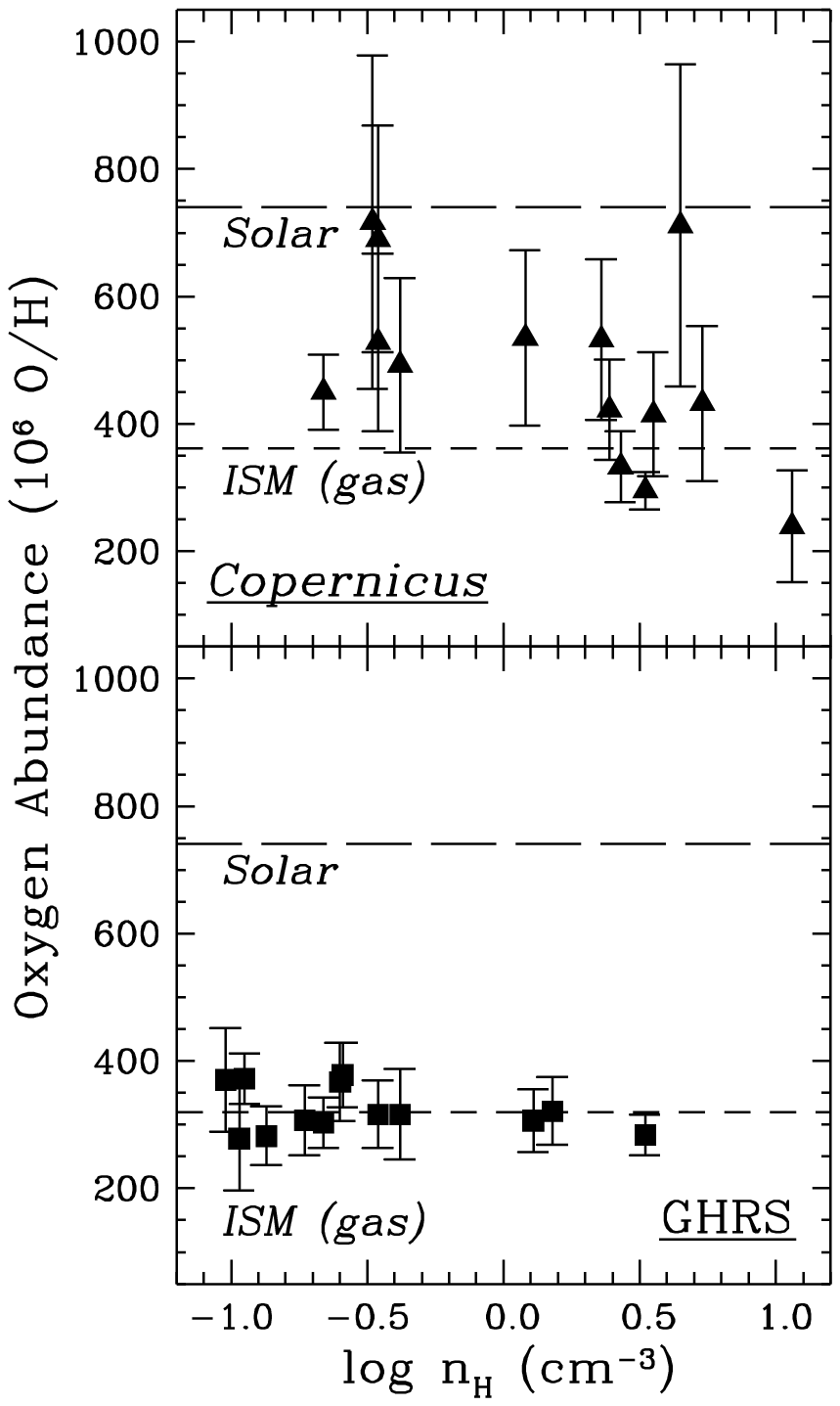]{The interstellar oxygen abundances measured in our
{\it Copernicus} and GHRS samples as a function of the logarithmic
mean hydrogen density, $n_H$ $=$ $N$(H)/$r$,
in the observed sightlines.  The short-dashed lines through the
data points represent the respective {\it Copernicus} and GHRS
weighted mean values (per 10$^6$ H
atoms) of 10$^6$ O/H $=$ 361 $\pm$ 20 and
10$^6$ O/H $=$ 319 $\pm$ 14 for the interstellar gas-phase
oxygen abundance.  The long-dashed lines represent the
Grevesse \& Noels (1993) solar oxygen abundance
(10$^6$ O/H $=$ 741 $\pm$ 130).}

\figcaption[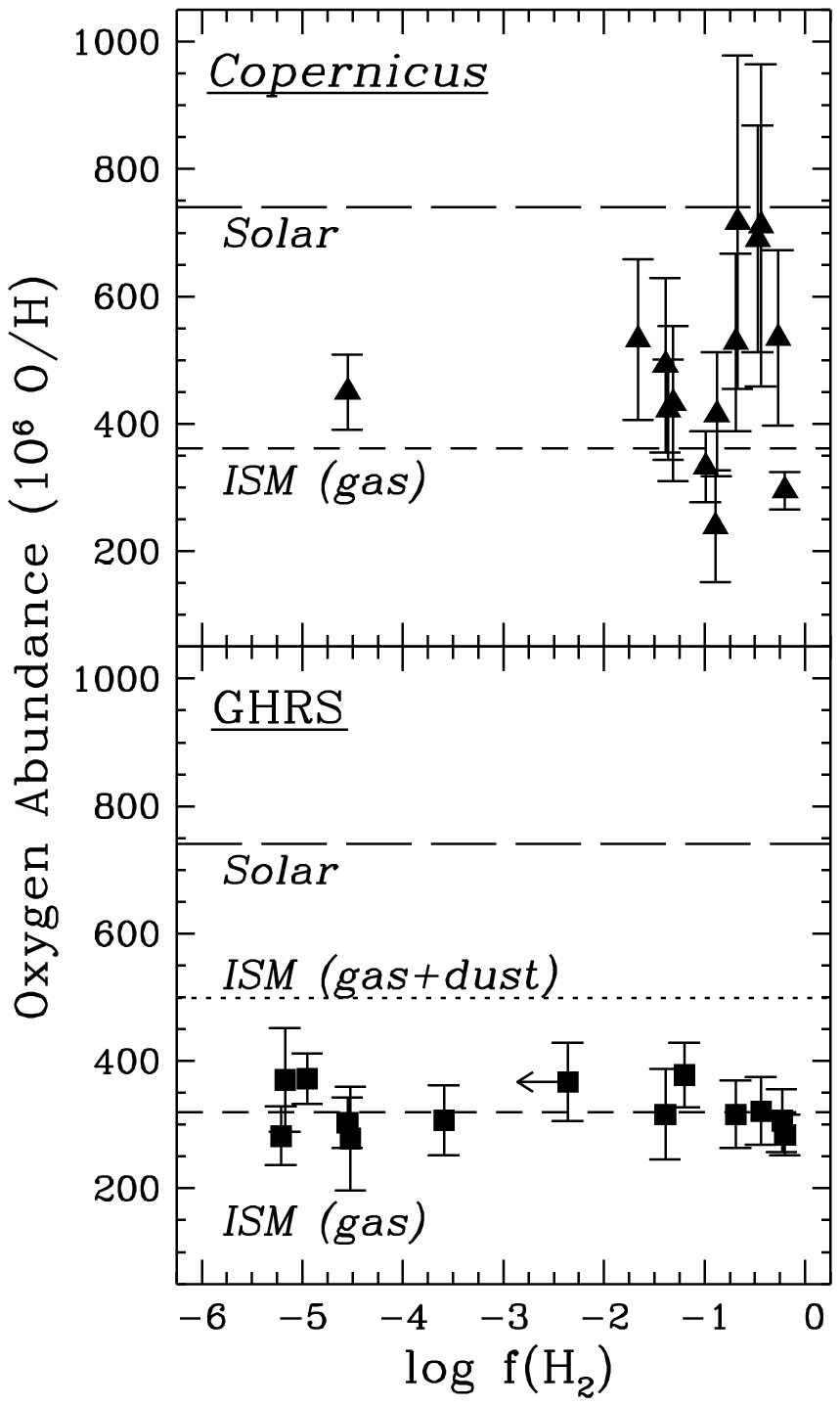]{The interstellar oxygen abundances measured in our
{\it Copernicus} and GHRS samples as a function of the logarithmic
fraction of hydrogen in molecular form,
$f$(H$_2$) $=$ 2$N$(H$_2$)/$N$(H),
in the observed sightlines.  The short-dashed lines through the
data points represent the respective {\it Copernicus} and GHRS
weighted mean values (per 10$^6$ H
atoms) of 10$^6$ O/H $=$ 361 $\pm$ 20 and
10$^6$ O/H $=$ 319 $\pm$ 14 for the interstellar gas-phase
oxygen abundance.  The dotted line above the GHRS data points
at 10$^6$ O/H $\approx$ 500 reflects the {\it total} (gas plus
dust) GHRS interstellar oxygen abundance after an allowance
is made for the O tied up in dust grains.  This value is about
2/3 of the Grevesse \& Noels (1993) solar oxygen abundance
(10$^6$ O/H $=$ 741 $\pm$ 130) denoted by the long-dashed lines
in the figure.}

\figcaption[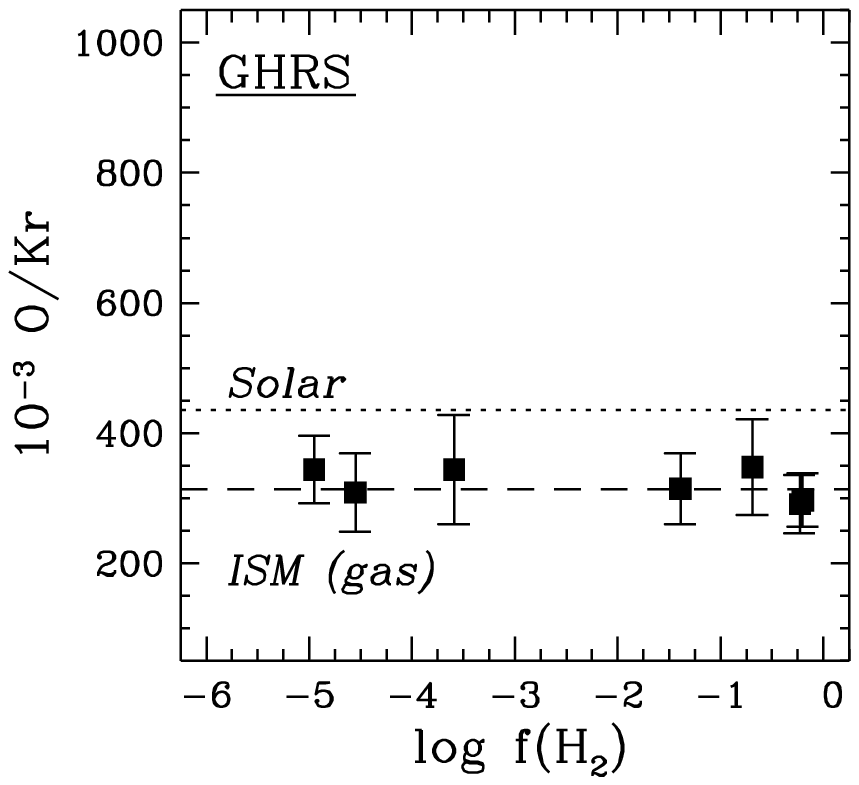]{The interstellar gas-phase O/Kr abundance ratioa
measured with GHRS as a function of the logarithmic
fraction of hydrogen in molecular form,
$f$(H$_2$) $=$ 2$N$(H$_2$)/$N$(H),
in the 7 observed sightlines with both O (this paper) and
Kr (Cardelli \& Meyer 1997) data.  The dashed line among the
data points reflects their weighted mean value of
10$^{-3}$ O/Kr $=$ 314 $\pm$ 20.  After accounting for the
O tied up in dust grains, this ratio comes close to the
solar value of 10$^{-3}$ O/Kr $=$ 436 $\pm$ 108 (Anders \&
Grevesse 1989, Grevesse \& Noels 1993).}

\clearpage

\begin{deluxetable}{lcccccc}
\footnotesize
\tablecaption{The GHRS Interstellar Oxygen Abundances \label{tbl1}}
\tablewidth{0pt}
\tablehead{
\colhead{Star} & \colhead{$N$(H)\tablenotemark{a}} &
\colhead{log $n_H$\tablenotemark{b}} &
\colhead{log $f$(H$_2$)\tablenotemark{c}} &
\colhead{$W_\lambda$(1356)\tablenotemark{d}} &
\colhead{$N$(O I)\tablenotemark{e}} &
\colhead{10$^6$ O/H\tablenotemark{f}}\\
\colhead{} & \colhead{(cm$^{-2}$)} & \colhead{(cm$^{-3}$)} &
\colhead{} & \colhead{(m\AA)} & \colhead{(cm$^{-2}$)} & \colhead{}}
\startdata
$\gamma$ Cas & 1.5 (0.2) $\times$ 10$^{20}$ & $-$0.60 & $<-$2.36
& \phn1.1 (0.1) & 5.4 (0.5) $\times$ 10$^{16}$ & 367 (62) \nl
$\zeta$ Per & 1.6 (0.2) $\times$ 10$^{21}$ & \phs0.11 & $-$0.23
& \phn8.0 (0.5) & 4.8 (0.6) $\times$ 10$^{17}$ & 306 (49) \nl
$\epsilon$ Per & 3.3 (0.5) $\times$ 10$^{20}$ & $-$0.46 & $-$0.69
& \phn2.1 (0.2) & 1.0 (0.1) $\times$ 10$^{17}$ & 316 (53) \nl
$\xi$ Per & 1.9 (0.2) $\times$ 10$^{21}$ & \phs0.18 & $-$0.44
& 10.8 (1.3) & 6.0 (0.8) $\times$ 10$^{17}$ & 321 (53) \nl
$\delta$ Ori & 1.6 (0.2) $\times$ 10$^{20}$ & $-$0.87 & $-$5.21
& \phn0.9 (0.1) & 4.4 (0.5) $\times$ 10$^{16}$ & 282 (46) \nl
$\lambda$ Ori & 6.5 (1.2) $\times$ 10$^{20}$ & $-$0.38 & $-$1.39
& \phn4.0 (0.5) & 2.0 (0.3) $\times$ 10$^{17}$ & 316 (71) \nl
$\iota$ Ori & 1.5 (0.2) $\times$ 10$^{20}$ & $-$1.02 & $-$5.17
& \phn1.1 (0.2) & 5.4 (1.0) $\times$ 10$^{16}$ & 370 (81) \nl
$\epsilon$ Ori & 2.9 (0.4) $\times$ 10$^{20}$ & $-$0.73 & $-$3.59
& \phn1.8 (0.2) & 8.9 (1.0) $\times$ 10$^{16}$ & 307 (55) \nl
$\kappa$ Ori & 3.4 (0.3) $\times$ 10$^{20}$ & $-$0.66 & $-$4.55
& \phn2.1 (0.2) & 1.0 (0.1) $\times$ 10$^{17}$ & 303 (40) \nl
15 Mon & 2.3 (0.4) $\times$ 10$^{20}$ & $-$0.97 & $-$4.52
& \phn1.3 (0.3) & 6.4 (1.5) $\times$ 10$^{16}$ & 278 (81) \nl
$\tau$ CMa & 5.3 (0.4) $\times$ 10$^{20}$ & $-$0.95 & $-$4.95
& \phn4.0 (0.3) & 2.0 (0.2) $\times$ 10$^{17}$ & 372 (40) \nl
$\zeta$ Oph & 1.4 (0.1) $\times$ 10$^{21}$ & \phs0.52 & $-$0.20
& \phn6.4 (0.6) & 4.0 (0.4) $\times$ 10$^{17}$ & 284 (32) \nl
$\gamma$ Ara & 5.4 (0.6) $\times$ 10$^{20}$ & $-$0.59 & $-$1.20
& \phn3.9 (0.2) & 2.0 (0.2) $\times$ 10$^{17}$ & 378 (51) \nl
\enddata
\tablenotetext{a}{$N$(H) $=$ 2$N$(H$_2$) $+$ $N$(H I) is the total
hydrogen column density ($\pm$1$\sigma$) in the observed sightlines.
These values reflect the H$_2$ column densities measured by
\cite{sav77} and the weighted means of the
\cite{boh78} and \cite{dip94} $N$(H I) data.}
\tablenotetext{b}{The mean hydrogen sightline density is calculated
from $N$(H) and the stellar distances.}
\tablenotetext{c}{$f$(H$_2$) $=$ 2$N$(H$_2$)/$N$(H) is the fractional
abundance of hydrogen nuclei in H$_2$ in the observed sightlines.}
\tablenotetext{d}{The measured equivalent widths ($\pm$1$\sigma$) of
the O I 1355.598 \AA\ absorption line.}
\tablenotetext{e}{The derived O I column densities ($\pm$1$\sigma$)
in the observed sightlines.  The $\xi$ Per and $\zeta$ Oph values are
taken from the analyses of \cite{car91} and \cite{sav92}.  The
$\zeta$ Per, $\lambda$ Ori, and $\gamma$ Ara values are corrected for
a slight amount of saturation using respective Gaussian
$b$-values ($\pm$1$\sigma$) of 2.0$^{+2.0}_{-0.5}$,
5.0$^{+\infty}_{-2.5}$, and 3.0$^{+\infty}_{-1.5}$
km s$^{-1}$.  The other sightlines are assumed to be optically thin
in the O I $\lambda$1356 transition.}
\tablenotetext{f}{The abundance of interstellar gas-phase oxygen
($\pm$1$\sigma$) per 10$^6$ H atoms in the observed sightlines.  The
uncertainties reflect the propagated $N$(H) and $N$(O I) errors.}
\end{deluxetable}

\clearpage

\begin{deluxetable}{lccccccc}
\footnotesize
\tablecaption{The {\it Copernicus} Interstellar Oxygen Abundances
\label{tbl2}}
\tablewidth{0pt}
\tablehead{
\colhead{Star} & \colhead{$N$(H)\tablenotemark{a}} &
\colhead{log $n_H$\tablenotemark{b}} &
\colhead{log $f$(H$_2$)\tablenotemark{c}} &
\colhead{$W_\lambda$(1356)\tablenotemark{d}} &
\colhead{$b$\tablenotemark{e}} &
\colhead{$N$(O I)\tablenotemark{f}} &
\colhead{10$^6$ O/H\tablenotemark{g}}\\
\colhead{} & \colhead{(cm$^{-2}$)} & \colhead{(cm$^{-3}$)} &
\colhead{} & \colhead{(m\AA)} & \colhead{(km s$^{-1}$)} &
\colhead{(cm$^{-2}$)} & \colhead{}}
\startdata
o Per & 1.5 (0.2) $\times$ 10$^{21}$ & \phs0.08 & $-$0.27
& 11.7 (2.4) & 2.0$^{+2.0}_{-0.5}$ & 8.0 (1.8) $\times$ 10$^{17}$
& 535 (138) \nl
$\epsilon$ Per & 3.3 (0.5) $\times$ 10$^{20}$ & $-$0.46 & $-$0.69
& \phn3.5 (0.8) & \nodata & 1.7 (0.4) $\times$ 10$^{17}$
& 528 (140) \nl
$\alpha$ Cam & 1.3 (0.1) $\times$ 10$^{21}$ & $-$0.46 & $-$0.47
& 15.5 (3.3) & 5.0$^{+\infty}_{-2.5}$ & 8.9 (2.1) $\times$ 10$^{17}$
& 690 (178) \nl
$\lambda$ Ori & 6.5 (1.2) $\times$ 10$^{20}$ & $-$0.38 & $-$1.39
& \phn6.1 (1.3) & 5.0$^{+\infty}_{-2.5}$ & 3.2 (0.7) $\times$ 10$^{17}$
& 492 (137) \nl
$\kappa$ Ori & 3.4 (0.3) $\times$ 10$^{20}$ & $-$0.66 & $-$4.55
& \phn3.1 (0.3) & \nodata & 1.5 (0.2) $\times$ 10$^{17}$
& 450 (\phn59) \nl
1 Sco & 1.6 (0.2) $\times$ 10$^{21}$ & \phs0.36 & $-$1.66
& 13.3 (2.5) & 3.0$^{+3.0}_{-1.0}$ & 8.3 (1.7) $\times$ 10$^{17}$
& 532 (126) \nl
$\delta$ Sco & 1.2 (0.2) $\times$ 10$^{21}$ & \phs0.39 & $-$1.36
& \phn9.2 (0.8) & 4.0$^{+\infty}_{-2.0}$ & 5.1 (0.7) $\times$ 10$^{17}$
& 422 (\phn79) \nl
$\beta^1$ Sco & 1.3 (0.1) $\times$ 10$^{21}$ & \phs0.43 & $-$0.99
& \phn7.8 (0.8) & 3.0$^{+\infty}_{-1.0}$ & 4.4 (0.7) $\times$ 10$^{17}$
& 333 (\phn56) \nl
$\omega^1$ Sco & 1.7 (0.3) $\times$ 10$^{21}$ & \phs0.55 & $-$0.88
& 11.9 (1.7) & 3.0$^{+3.0}_{-1.0}$ & 7.2 (1.1) $\times$ 10$^{17}$
& 415 (\phn97) \nl
$\sigma$ Sco & 2.5 (0.3) $\times$ 10$^{21}$ & \phs0.73 & $-$1.31
& 14.3 (1.6) & 2.0$^{+2.0}_{-0.5}$ & 1.1 (0.3) $\times$ 10$^{18}$
& 432 (122) \nl
$\rho$ Oph & 5.7 (0.7) $\times$ 10$^{21}$ & \phs1.06 & $-$0.89
& 16.5 (3.3) & 2.0$^{+2.0}_{-0.5}$ & 1.4 (0.5) $\times$ 10$^{18}$
& 239 (\phn88) \nl
$\chi$ Oph & 2.4 (0.3) $\times$ 10$^{21}$ & \phs0.65 & $-$0.44
& 26.0 (6.0) & 5.0$^{+\infty}_{-2.0}$ & 1.7 (0.6) $\times$ 10$^{18}$
& 712 (253) \nl
$\zeta$ Oph & 1.4 (0.1) $\times$ 10$^{21}$ & \phs0.52 & $-$0.20
& \phn6.6 (0.4) & 1.5$^{+0.5}_{-0.3}$ & 4.1 (0.3) $\times$ 10$^{17}$
& 295 (\phn29) \nl
15 Sgr & 1.8 (0.3) $\times$ 10$^{21}$ & $-$0.48 & $-$0.67
& 20.8 (4.3) & 5.0$^{+\infty}_{-2.5}$ & 1.3 (0.4) $\times$ 10$^{18}$
& 717 (262) \nl
\enddata
\tablenotetext{a}{$N$(H) $=$ 2$N$(H$_2$) $+$ $N$(H I) is the total
hydrogen column density ($\pm$1$\sigma$) in the observed sightlines.
These values reflect the H$_2$ column densities measured by
\cite{sav77} and the weighted means of the
\cite{boh78} and \cite{dip94} $N$(H I) data.}
\tablenotetext{b}{The mean hydrogen sightline density is calculated
from $N$(H) and the stellar distances.}
\tablenotetext{c}{$f$(H$_2$) $=$ 2$N$(H$_2$)/$N$(H) is the fractional
abundance of hydrogen nuclei in H$_2$ in the observed sightlines.}
\tablenotetext{d}{The measured equivalent widths ($\pm$1$\sigma$) of
the O I 1355.598 \AA\ absorption line from \cite{boh83} and
\cite{zei77} ($\zeta$ Oph).}
\tablenotetext{e}{The Gaussian $b$-values ($\pm$1$\sigma$) used to
correct the O I $\lambda$1356 line for saturation in deriving
$N$(O I).  The sightlines toward $\epsilon$ Per and $\kappa$ Ori
are assumed to be optically thin in this transition.}
\tablenotetext{f}{The derived O I column densities ($\pm$1$\sigma$)
in the observed sightlines.}
\tablenotetext{g}{The abundance of interstellar gas-phase oxygen
($\pm$1$\sigma$) per 10$^6$ H atoms in the observed sightlines.  The
uncertainties reflect the propagated $N$(H) and $N$(O I) errors.}
\end{deluxetable}

\end{document}